\documentclass[preprint,aps]{revtex4}
\usepackage[latin1]{inputenc}
\usepackage{amsmath, fancyhdr}
\usepackage{scrtime}
\usepackage{epsfig}
\usepackage{floatflt}
\usepackage{color}
\newcommand{\comment}[1]{}
\newcommand{\intli}{\int\limits}
\newcommand{\bra}[3]{\left\langle #1 \left | #2 \right | #3 \right\rangle}
\newcommand{\phial}{\phi_{\alpha}}
\newcommand{\phibe}{\phi_{\beta}}
\newcommand{\braphi}[1]{\bra{\phial}{#1}{\phibe}}
\newcommand{\braa}[2]{\left\langle #1 \left | #2 \right . \right\rangle}
\newcommand{\brab}[2]{\left\langle\left . #1 \right | #2 \right\rangle}
\newcommand{\ddt}{\frac{d}{dt}}

\newcommand{\parfrac}[2]{\frac{\partial #1}{\partial #2}}
\newcommand{\delfrac}[2]{\frac{\delta #1}{\delta #2}}
\newcommand{\parfraca}[1]{\frac{\partial }{\partial #1}}
\newcommand{\R}{\mathbf R}
\newcommand{\E}{\mathbf E}
\newcommand{\B}{\mathbf B}
\newcommand{\C}{\mathbf C}

\newcommand{\RN}{\R}
\newcommand{\RAal}{\R_{A_{\alpha}}}
\renewcommand{\r}{\mathbf r}
\newcommand{\fracb}[1]{\frac{#1}{|\r-\r'|} }
\newcommand{\ifracb}[1]{\int{ \fracb{#1} d^3r'} }

\newcommand{\Ac}{A_{\rm c}}
\newcommand{\Vint}{V_{\rm int}}
\newcommand{\Ne}{N_{\rm e}}
\newcommand{\Ni}{N_{\rm i}}
\newcommand{\de}{ {\mathbf d}_{\rm e}}
\newcommand{\di}{ {\mathbf d}_{\rm i}}
\newcommand{\Apot}{A_{\rm pot}}
\newcommand{\Aq}{A_{\rm q}}
\newcommand{\Fq}{F_{\rm q}}
\newcommand{\Axc}{A_{\rm xc}}

\newcommand{\BAalbe}{\mathbf B^A_{\alpha\beta}}
\newcommand{\CAalbe}{\mathbf C^A_{\alpha\beta}}

\newcommand{\Bbega}{B_{\beta\gamma}}
\newcommand{\BAdebe}{\mathbf B^A_{\delta\beta}}
\newcommand{\BApalga}{\mathbf B^{A+}_{\alpha\gamma}}
\newcommand{\BApalbe}{\mathbf B^{A+}_{\alpha\beta}}
\newcommand{\CApalbe}{\mathbf C^{A+}_{\alpha\beta}}

\newcommand{\Bpalga}{B^+_{\alpha\gamma}}
\newcommand{\Bpalbe}{B^+_{\alpha\beta}}
\newcommand{\Bdebe}{B_{\delta\beta}}
\newcommand{\Balbe}{B_{\alpha\beta}}

\newcommand{\DAalbe}{\mathbf D^A_{\alpha\beta}}

\newcommand{\Exc}{E_{\rm xc}}
\newcommand{\Epot}{E_{\rm pot}}

\newcommand{\Halbe}{H_{\alpha\beta}}
\newcommand{\Halga}{H_{\alpha\gamma}}
\newcommand{\Hbega}{H_{\beta\gamma}}
\newcommand{\Hdebe}{H_{\delta\beta}}

\renewcommand{\P}{\mathbf P}
\newcommand{\Pc}{\mathbf P_{\rm c}}
\newcommand{\Pq}{\mathbf P_{\rm q}}
\newcommand{\Salbe}{S_{\alpha\beta}}
\newcommand{\Sigade}{S^{-1}_{\gamma\delta}}
\newcommand{\Sialbe}{S^{-1}_{\alpha\beta}}

\newcommand{\Talbe}{T_{\alpha\beta}}
\newcommand{\U}{U(\RN, t)}
\newcommand{\Veff}{V_{\rm eff}}
\newcommand{\Vxc}{V_{\rm xc}}
\newcommand{\VL}{V_{\rm L}}

\newcommand{\ajals}{a^{j*}_{\alpha}}
\newcommand{\ajal}{a^j_{\alpha}}
\newcommand{\ajbe}{a^j_{\beta}}

\newcommand{\ajga}{a^j_{\gamma}}
\newcommand{\inttt}{\intli_{t0}^{t1}}
\newcommand{\lpsi}{\left |\Psi \right\rangle}

\newcommand{\lbra}[1]{\left |#1 \right\rangle}
\newcommand{\rbra}[1]{\left \langle #1 \right|}
\newcommand{\pdRA}{\frac{\partial}{\partial \R_A}}
\newcommand{\pdt}[1]{\frac{\partial #1}{\partial t}}

\newcommand{\sumA}{\sum_{A=1}^{\Ni} }
\newcommand{\sumj}{\sum_{j=1}^{\Ne} }
\newcommand{\sumalbe}{\sum_{\alpha\beta}}

\newcommand{\sumbega}{\sum_{\beta\gamma}}
\newcommand{\sumjalbe}{\sumj \sum_{\alpha\beta}}

\newcommand{\sumgade}{\sum_{\gamma\delta}}
\newcommand{\Hp}{H$_2^+$}

\newcommand{\beqx}[1]{\begin{equation} #1 \end{equation}}
\newcommand{\xref}[1]{(\ref{#1})}
\newcommand{\labex}[1]{
        \label{#1}
}
\begin{document}
\title{ Non-adiabatic quantum molecular dynamics: \\ Generalized formalism including 
laser fields}

\author{Thomas Kunert}
% \email{kunert@physik.tu-dresden.de}
 
\author{Rüdiger Schmidt}
% \email{schmidt@physik.tu-dresden.de}  
\affiliation{
        Institut für Theoretische Physik \\
        Technische Universität Dresden, 01062 Dresden 
}

%\date{\today}

\begin{abstract}
A generalized formalism of the so-called non-adiabatic quantum
molecular dynamics is presented, which applies for atomic many-body
systems in external laser fields. The theory treats the nuclear
dynamics and electronic transitions simultaneously in a mixed
classical-quantum approach. Exact, self-consistent equations of motion
are derived from the action principle by combining time-dependent
density functional theory in basis expansion with classical molecular
dynamics.  Structure and properties of the resulting equations of
motion as well as the energy and momentum balance equations are
discussed in detail. Future applications of the formalism are briefly
outlined.
\end{abstract}

\maketitle

\section{Introduction}

The non-adiabatic dynamics of electronic and nuclear degrees of
freedom in atomic many-body systems represents one of the fundamental
processes in different areas of physics and chemistry.

Experimentally, exceptional large progress has been made during the
last decade in studying non-adiabatic processes, in particular in
molecules and atomic clusters. So, experiments with intense
femto-second laser pulses interacting with molecules \cite{Bandrauk94}
or atomic clusters
\cite{Ditmire96b,Ditmire97b,Pherson94,Lezius98,Ditmire99,Haberland98,Broer99}
have revealed a variety of fascinating new, typical non-adiabatic
phenomena like the production of keV electrons \cite{Ditmire96b}, MeV
ions \cite{Ditmire97b} and intense x-rays \cite{Pherson94}; the Coulomb
explosion \cite{Lezius98} connected even with nuclear fusion
\cite{Ditmire99}; the multiple plasmon excitation and relaxation in metallic
clusters \cite{Haberland98}, or the unexpected enhanced ionization
with decreasing laser intensity \cite{Broer99}.
Moreover, pump-probe experiments allow now to investigate the
time-resolved non-adiabatic dynamics, e.g. of photoinduced
isomerization processes (for a review see \cite{Domcke97}). Finally,
refined scattering experiments involving metal clusters
\cite{Brenot96} and fullerenes \cite{Opitz00} revealed 
detailed insight into electronic and vibronic excitation mechanisms,
as well as their coupling and related fragmentation processes in those
complex systems.

Theoretically, the non-adiabatic coupling of electronic and nuclear
dynamics is one of the most challenging problems of atomic many-body
theory and, in principle, requires the solution of the full
time-dependent electron-nuclear Schrödinger equation. At present,
however, a full-scale numerical solution is barely feasible for the
smallest possible molecular system, the \Hp molecule
\cite{Bandrauk95}. Thus, for larger systems like atomic clusters,
phenomenological models, based on classical mechanics and/or
hydrodynamics
\cite{Petruck97,Last99,Ditmire98a,Blenski00,Boyer94,Rost01} have been
developed to investigate the mechanism of the intense laser-cluster
interaction. More microscopic approaches are based on electronic
time-dependent Thomas-Fermi theory
\cite{Brewczyk98,Brewczyk99,Blenski01} or related semiclassical
(meanfield) approximations \cite{Gattuta98,Garcia01} coupled with
molecular dynamics (MD) for the nuclear motion.  The most advanced
microscopic theory to study the coupled electronic and ionic dynamics
in intense laser-cluster interaction developed so far, is based on
time-dependent (TD) density functional theory (DFT) in local density
approximation (LDA) for the treatment of the electronic system coupled
with classical MD for the nuclear (ionic) dynamics
\cite{Reinhard98a,Reinhard00b}.  In this approach, the TD-Kohn-Sham
equations are numerically solved on a grid with the consequence that
full 3D calculations \cite{Reinhard98a} are still on the edge of
available computational facilities. Therefore, the upper most
applications of this theory have been obtained within an effective
two-dimensional approximation \cite{Reinhard00b} (see \cite{Reinhard00} for a
review).

An alternative fully microscopic approach to the nonadiabatic dynamics
in atomic many-body systems is the so-called nonadiabatic quantum
molecular dynamics (NA-QMD), developed recently \cite{Saalmann96}. In
this method, electronic and vibrational degrees of freedom are treated
simultaneously and self-consistently by combining classical MD with
TD-DFT in a finite-basis-set expansion of the Kohn-Sham-orbitals. The
formalism \cite{Saalmann96} has been worked out for conservative
systems, in particular to investigate adiabatic and non-adiabatic
collisions involving molecules and atomic clusters. So the NA-QMD
theory has been successfully applied so far for the description and
interpretation of fragment correlations in collision-induced
dissociation \cite{Fayeton98}, charge transfer cross sections
\cite{Knospe99b,Knospe99,Roller99}, as well as the excitation and 
fragmentation processes \cite{Saalmann98,Kunert01} in collisions of atoms (ions) with small
sodium clusters  and systems as large as fullerenes.

In this work, we present a generalization of the NA-QMD formalism
\cite{Saalmann96} (hereafter refered to as I), suitable to describe also the interaction of large,
but still finite atomic many-body systems with external laser
fields. We derive and discuss the exact equations of motion in a
systematic way, starting from a general mixed classical-quantum
treatment. Energy and momentum balance equations are derived as
well. Necessary approximations and possible simplifications to the
exact equations of motion as well as future applications of the
formalism are briefly summarized.

\section{Theory}
\subsection{General mixed classical-quantum treatment}

We consider first the general case of a mixed classical-quantum system
consisting of interacting particles. The $\Ni$ classical particles are
described by their trajectories $\R\equiv\lbrace\R_1(t),\dots,
\R_{\Ni}(t) \rbrace$.  They are determined by an explicit
time-dependent potential $U(\R,t)$ as well as the interaction with a
system of $\Ne$ quantummechanical particles, described by their
many-body wave function $\Psi=\Psi( \r_1, \dots, \r_{\Ne}, t)$ 
(We omit the spin index).
This is determined by an explicit time-dependent Hamiltonian $\hat
H(\R, t)$ which on his part depends parametrically on $\R$. The action
of such a system consists of a classical and a quantum part
\beqx{ \labex{Action0}A=\Ac+\Aq }%finish
with
\beqx{ 
        \labex{Ac0}
        \Ac=\inttt \left\lbrace \sum_A^{\Ni}{ \frac {M_A} 2 \dot \R_A^2} - \U \right\rbrace dt 
}%finish
and (atomic units $\hbar=e=m_e=\frac{1}{4\pi\varepsilon_0}=1$ are used)
\beqx{ 
        \labex{Aq0}
        \Aq=\inttt \bra{\Psi}{ i \pdt{}-\hat H(\RN, t )}{ \Psi } dt 
}%finish
with $M_A$ the mass of the classical particles and the brackets
$\langle\dots\rangle$ denote integration over all coordinates $\r_1,
\dots, \r_{\Ne}$. The equations of motion for the trajectories $\R$ and the many body state 
$| \Psi \rangle$ are obtained by making the total action stationary, leading to

\begin{alignat}{2}
  \labex{Schro0}
        \delfrac{A}{\langle\Psi(t)| }&=0  \Rightarrow \quad & i \pdt{} \lpsi &= \hat H(\RN, t)\lpsi  \\
   \labex{Clas0}
        \delfrac{A}{\R_A(t) }&=0  \Rightarrow \quad & M_A\ddot\R_A
        &=-\parfraca{\R_A}U(\R, t)-\bra{\Psi}{\pdRA \hat H(\R, t)}{\Psi} \\
             \notag   &&A&=1,\dots,\Ni \quad.
\end{alignat}
Equations (\ref{Schro0}) and (\ref{Clas0}) have to be
solved simultaneously. They represent the general equations of motion of
the mixed classical-quantum system defined above. They are much more universal than
those derived in I from energy conservation. Here they are obtained from a general 
action principle where both, the potential $U$ (defining the classical
system) and the Hamiltonian $\hat H$ (defining the quantum system as
well as the coupling to the classical one) may explicitely depend on
time. There is no energy or momentum conservation, nevertheless
classical motion $\R(t)$ and quantum dynamics $|\Psi(t)\rangle$ are coupled
self-consistently owing to the action principle.

In the next subsection, the potential U and the Hamiltonian $\hat H$
will be specified for an atomic many body system, we are interested in.

\subsection{Atomic many body system}

Considering now $\Ni$ ions (nuclei) with charge $Z_A \; (A=1, \dots,\Ni)$
and $\Ne$ electrons exposed to an external laser potential (usually,
but not necessarily, described in dipole approximation 
$V_{\rm L}(\mathbf x, t)=-\mathbf x\cdot\mathbf E(t)$, with $\mathbf E(t)$ the electric field strength)
the potential energy of the nuclei reads

\beqx{ 
        \labex{U}
        \U = \sum_{A<B}^{\Ni}\frac{Z_A Z_B}{| \R_A-\R_B |}-\sum_{A=1}^{\Ni} Z_A \VL(\R_A, t) 
}%finish
and the Hamiltonian becomes
\beqx{
        \labex{Hamil1}  
        \hat H(\RN, t)= \sum_{i=1}^{\Ne} \hat t_i + \sum_{i=1}^{\Ne} V(\r_i, \R, t)
        +\sum_{i<j}^{\Ne}\frac{1}{|\r_i-\r_j|} 
}%finish
with the single particle kinetic energy operator $\hat
t=-\frac\Delta 2$.  The external single particle potential $V(\r, \R, t)$
contains the laser field $\VL(\r, t)$ and the electron-ion interaction $\Vint(\r, \R )$
        \stepcounter{equation}
        \newcounter{vadummy}
        \setcounter{vadummy}{\value{equation}}

\begin{align}
   \labex{V1a}
        \tag{ \arabic{equation}a}
        V( \r, \R, t ) &= \Vint( \r, \R ) + \VL(\r, t) \\
   \labex{V1b}
        \tag{ \arabic{equation}b}
                        &=-\sum_{A=1}^{\Ni}\frac{Z_A}{|\R_A-\r|}+\VL(\r, t) \quad .
\end{align}
The first term in (\arabic{vadummy}) is time dependent via $\R(t)$ 
and the second one explicitely depends on time. Using the definition of the single particle density
\beqx{
        \labex{Dens0}
        \rho(\r, t) = \Ne\cdot\int d^3 r_2 \dots d^3r_{\Ni} \Psi^*(\r, \r_2, \dots,\r_{\Ne}, t) \cdot
        \Psi(\r, \r_2, \dots, \r_{\Ne}, t)
} % finish
it becomes apparent that the general Newton-type equation (\ref{Clas0}) drastically simplifies
with (\ref{Hamil1}), (\arabic{vadummy}), (\ref{Dens0}) leading to
\beqx{
        \labex{Clas1}
        M_A\ddot\R_A = - \parfrac{U(\R,t)}{\R_A}
        - \int d³\r \rho(\r,t) \parfrac{\Vint(\r,\R)} {\R_A} 
             \qquad   A=1,\dots,\Ni \quad .
}
Thus, the electronic forces acting on the nuclei are determined by the single particle 
density $\rho(\r, t)$ 
alone, which is the key quantity in DFT. So, in the next subsection we will reformulate the whole problem using TD-DFT to describe the electronic system.

\subsection{TD-DFT coupled with MD}

According to the basic theorems of TD-DFT \cite{Gross84} any
observable of a many body system can be expressed as functional of the
single particle density (\ref{Dens0}) and this density can be
obtained from a non-interacting reference system according to the ansatz
\beqx {
        \labex{Rho1}
        \rho(\r,t)=\sumj\Psi^{j*}(\r,t) \Psi^j(\r,t)
}
with $\Psi^j(\r,t)$ the time dependent Kohn Sham functions.
The quantum mechanical part of the action (\ref{Action0}) now reads
\beqx{
        \labex{Aq1}
        \Aq = \inttt\sumj\bra{\Psi^j}{i\pdt{}-\hat t}{\Psi^j}dt-\Apot
}
where the brackets $\langle \dots \rangle \equiv\int_V d^3r$ denote integration
over the single particle coordinate. The potential part in (\ref{Aq1})
\beqx{
        \labex{Apot1}
        \Apot=\inttt\int\rho(\r, t)\Bigg(V(\r, \R, t)+
        \frac12\ifracb{\rho(\r', t)}\Bigg)d^3r \,dt+\Axc[\rho]
}
 is a functional of the density $\rho(\r,t)$ and contains the
 exchange-correlation contribution $\Axc$. In concrete applications of
 TD-DFT, the latter is subject of adequate approximations, like the
 time dependent local density approximation (TD-LDA) or the time
 dependent optimized potential method \cite{Gross98}. In this paper we will not specify $\Axc$ 
and, thus, are dealing with exact equations of motion.

In this sense, variation of (\ref{Aq1}), (\ref{Apot1}) with respect to the KS-orbitals leads to
\beqx{  
        \labex{KS1} 
        \delfrac{A}{\Psi^{j*}(\r,t) }=0  
        \Rightarrow \quad  i \pdt{} \Psi^j = (\hat t + \Veff(\r, \RN, t))\Psi^j,\quad  j=1,\dots,\Ne
}%finish
whereas, variation of (\ref{Ac0}), (\ref{Aq1}) and (\ref{Apot1})
with respect to the trajectories gives

\beqx{
\begin{split}
        \labex{Clas2} 
        \delfrac{A}{\R_A(t) }=0 \Rightarrow \quad
        &M_A\ddot\R_A=-\parfraca{\R_A}U(\R,t)
        -\sumj\bra{\Psi^j}{\pdRA \Vint(\r, \R)}{\Psi^j} \\
        &A=1,\dots,\Ni 
\end{split}
}
In (\ref{KS1}), the effective single particle potential $\Veff(\r, \R, t)$ is
defined as

\beqx{
\begin{split}
        \labex{Veff}
        \Veff(\r, \R, t) &= \frac{\delta \Apot[\rho]}{\delta \rho(\r, t)} \\
        &=V(\r, \R, t)+\ifracb{\rho(\r', t)}+\frac{\delta\Axc[\rho]}{\delta\rho(\r, t)} \quad .
\end{split}
}%finish
In (\ref{Clas2}), the interaction potential $\Vint(\r, \R)$, as part of $V(\r, \R, t)$, is 
defined according to (\arabic{vadummy}).

The resulting equations of motion (\ref{KS1}), (\ref{Clas2}) are completely equivalent to 
(\ref{Schro0}), (\ref{Clas0}) and accordingly to (\ref{Clas1}).
So, with the help of (\ref{Rho1}) one immediately realizes that
(\ref{Clas2}) is identical to (\ref{Clas1}). The many body
Schrödinger equation (\ref{Schro0}), however, is now replaced by a
set of $\Ne$ coupled integro-differential single particle KS-equations
(\ref{KS1}). In the present form, these equations have to be solved
numerically on a grid, which still is very demanding (if not impossible, at
present, for large systems in intense laser fields; see also
discussion in the next section.). A drastic simplification can be
achieved, if the (3+1)-dimensional KS-orbitals $\Psi^j(\r,t)$ are
represented in a finite basis set. This, however, complicates the
classical equations of motion (\ref{Clas2}) considerably as will be
discussed in the next section.

\subsection{TD-DFT in basis expansion coupled with MD}

In this section, we derive the final equations of motion of the general NA-QMD formalism
and discuss their properties, in particular the resulting energy and momentum balance equations.

The central starting point is to expand the time dependent KS-orbitals $\Psi^j (\r,t)$ in a local
basis $\lbrace\phi_\alpha\rbrace$
\beqx{
        \labex{Basis}
        \Psi^j(\r,t)=\sum_\alpha \ajal(t)\phi_\alpha(\r-\R_{A_\alpha})
} %finish
with the expansion coefficients $\ajal(t)$ and the symbol $A_\alpha$ denotes the atom to 
which the atomic orbital $\phial$ is attached.

Although technical details are not the topic of this paper, we note in
passing, that the use of the linear combination of atomic orbitals
(LCAO-ansatz (\ref{Basis})) has clear advantages as compared to a
direct numerical solution of the Kohn-Sham equations (\ref{KS1}).
First of all (and obviously), the (3+1)-dimensional problem(\ref{KS1})) will be
reduced to a one-dimensional one for the determination of the
coefficients $\ajal(t)$. Second (and especially important, if intense
laser fields are considered), electrons with basically different
spatial extensions (strongly bound core electrons, binding valence
electrons as well as practically free electrons in the continuum) can
be naturally included in the dynamical treatment, provided appropriate 
basis functions $\phial$ are taken into account \cite{tobe}.

With the ansatz (\ref{Basis}) the explicit expression of the density is given by

\beqx{ \labex{Dens} 
        \rho(\r,t)=\sumj\sumalbe\ajals(t)\ajbe(t)\phial^*(\r-\R_{A_\alpha})\phibe(\r-\R_{A_\beta})\quad.
}%finish

Owing to the implicit time-dependence of the basis
$\phial(\r-\R_{A_\alpha})$, the partial time derivative $\pdt{}$ in
the action (\ref{Aq1}) has to be replaced by
\beqx{
        \labex{pdt}
        \pdt{} \quad \Rightarrow \quad \ddt=\pdt{}+\sumA\dot\R_A\pdRA
} %finish
For the following considerations it is convenient to introduce the following matrices: \\
the kinetic energy matrix
\beqx{ 
        \Talbe:=\braphi{\hat t} \quad, 
}%finish
the hamilton matrix
\beqx{ 
        \labex{Halbe}
        \Halbe:=\braphi{\hat t+\Veff} \quad
}%finish
containing the effective potential $\Veff$ defined in (\ref{Veff}),
the overlap matrix
\beqx{ 
        \Salbe:=\braa{\phial}{\phibe} \quad , 
}%finish
the non-adiabatic coupling matrices
\beqx{ 
        \Balbe:=\braa{\phial}{\ddt\phibe}%=\dot\R_A\BAalbe 
}%finish
which due to (\ref{pdt}) contains the vector matrices
\beqx{
        \BAalbe:=\braa{\phial}{\pdRA\phibe} \quad, 
}%finish
and finally, the double differential matrix
\beqx{
        \CAalbe := \braa{\ddt\phial}{\pdRA\phibe}\quad.
}%finish
In addition, we define the transposed matrices
\begin{alignat}{2}
        \Bpalbe &:= \brab{\ddt\phial}{\phibe}&&=B_{\beta\alpha}^{*} \quad\\
        \BApalbe &:= \brab{\pdRA\phial}{\phibe}&&=\B_{\beta\alpha}^{A*} \quad \\
        \CApalbe &:= \brab{\pdRA\phial}{\ddt\phibe}&&=\C_{\beta\alpha}^{A*} \quad. 
\end{alignat}

With these definitions and the ansatz (\ref{Basis}) the quantum mechanical action 
(\ref{Aq1}) can be written as

\beqx{ \Aq=\inttt \Fq(t) dt-\Apot  }
with
\beqx{ 
        F_q(t)=\sumjalbe\ajals\left[
        (i\Balbe-\Talbe)\ajbe+i\Salbe\dot\ajbe
        \right ] \quad .
}%finish

The final equations of motion are now obtained by independent
variation of the total action with respect to $\ajal(t)$ and
$\R_A(t)$. With 
\beqx{
        \delfrac{A}{\ajals(t)}=\parfrac{\Fq}{\ajals}
        -\int d^3r \parfrac{\rho}{\ajals}\delfrac{\Apot}{\rho(\r,t)} = 0
}
this yields the Kohn-Sham equations in basis representation
\beqx{
        \labex{KS}
        \dot\ajal=-\sumbega\Salbe^{-1}\left(i\Hbega+\Bbega\right)\ajga\qquad j=1,\dots,\Ne
}%finish
and using Euler's equations
\beqx{ 
        \delfrac{A}{\R_A(t)} =
        \parfrac{F_q}{\R_A} - \ddt \parfrac{F_q}{\dot\R_A} 
        -\delfrac{\Apot}{\R_A(t)}
        +\delfrac{\Ac}{\R_A(t)} = 0 
}%finish
one obtains after some algebra the classical equations of motion
\beqx{ 
        \labex{Clas} M_A \ddot\R_A=-\parfrac{U(\R,t)}{\R_A}
        +\sumjalbe\ajals\left(-\parfrac{\Halbe}{\R_A}+\DAalbe\right)\ajbe \qquad A=1,\dots,\Ni
}%finish
with the matrix
\begin{multline}
        \labex{D}
  \DAalbe=\braphi{\pdRA(\Veff-V)} 
  +\sumgade\left(\BApalga\Sigade\Hdebe+\Halga\Sigade\BAdebe\right) \\
 +i\left[\CApalbe-\CAalbe
  +\sumgade\left(\Bpalga\Sigade\BAdebe-\BApalga\Sigade\Bdebe\right)\right] \quad .
\end{multline}

Equations \xref{KS} and (\ref{Clas}) represent the central result in
the derivation of the generalized formalism of the NA-QMD. The
Kohn-Sham equations (\ref{KS}) are formally very similar to that
derived in I, but now also contain the laser field which couples to
the electronic system (see definition $\Halbe$ \xref{Halbe}, $\Veff$
\xref{Veff}, $V$(\arabic{vadummy}) ). The classical equations
\xref{Clas} include the laser field as well, which here acts on the
nuclei (see definition of $\U$ \xref{U} ). Moreover, the quantum part
of the forces in \xref{Clas} differs appreciable from that derived in
I obtained from energy conservation. In particular, the last term in 
\xref{D} results from the variational principle. It represents an 
important contribution if the momentum balance is considered in the
present basis set formalism (see below). Obviously, this term
vanishes if the basis is complete, i.e. if
\beqx{
        \labex{Complete}
        \sumalbe \lbra{\phial} \Sialbe \rbra{\phibe}=1 \quad 
}
holds. It will be shown below, that in this case also the remaining terms
of the electronic contribution to the forces in \xref{Clas}, \xref{D}
are drastically simplified. In any practical applications of the formalism, however, the 
completeness relation \xref{Complete} can never be fulfilled, and thus, the full equations 
of motion \xref{Clas} have to be considered. 

At first glance, the complicated structure of the forces in
\xref{Clas}, \xref{D} makes it difficult to give a transparent
interpretation of the correction term resulting from
the basis. From the theoretical point of view it is therefore very useful to present
\xref{Clas}, \xref{D} in an alternative (operator) form and rederive the KS-equations \xref{KS} from a {\em basis constrained} single particle hamiltonian defined as
\beqx{
        \hat h' = \hat t + \Veff + \hat X
}
with $\hat h=\hat t+\Veff$ the usual KS-hamiltonian from \xref{KS1},
\xref{Veff} and the additional operator
\beqx{
        \labex{X}
        \hat X := \hat P\hat h\hat P-\hat h+i(1-\hat P)\hat B -i \hat B^+(1-\hat P) 
}
defined with the projectors
\beqx{
        \labex{P}
         \hat P := \sumalbe \lbra{\phial}\Sialbe\rbra{\phibe} 
}
and
\beqx{ 
        \labex{B}
        \hat B := \sumalbe \lbra{\ddt\phial}\Sialbe\rbra{\phibe}\quad .  
}
Obviously $\hat X$ vanishes for a complete basis \xref{Complete}.

With \xref{X}, \xref{P}, \xref{B} the classical equations of motion \xref{Clas} can now 
be rewritten as
\beqx{ 
    \labex{Clas3} 
    M_A\ddot\R_A=-\parfrac{U(\R,t)}{\R_A}-\sumjalbe\ajals
    \bra{\phial}{\pdRA \Vint(\r,\R) + \pdRA \hat X}{\phibe} 
    \ajbe
}
leading finally, with \xref{Basis}, to
\beqx{ 
  \labex{Clas4} 
  M_A\ddot\R_A=-\parfrac{U(\R, t)}{\R_A}-\sum_j\bra{\Psi^j}{\pdRA\Vint(\r, \R)+\pdRA\hat X}{\Psi^j}
   \quad . 
}
In addition, the equations of motion \xref{KS} are equivalent to the standard form of the 
time-dependent KS-equations
\beqx{ 
        \labex{KS4} 
        i\pdt{} \Psi^j=(\hat t + \Veff(\r, \R, t)+\hat X)  \Psi^j 
}
however, with the additional single particle operator $\hat X$ \xref{X}. This can easily be seen by inserting the ansatz \xref{Basis} into \xref{KS4} which leads to
\beqx{
        \sum_{\alpha} \left [ \dot\ajal + \sumbega\Sialbe(i\Hbega+\Bbega)\ajga\right ]\phial=0 
}
and, therefore, finally to \xref{KS} because the basis $\lbrace \phial
\rbrace$ must be linearly independent.

The implicit equations of motion \xref{Clas4} and \xref{KS4} are thus
completely equivalent to the explicit expressions \xref{KS},
\xref{Clas}, used in practical calculations. They allow however a more transparent 
interpretation of the present theory: \\ The use of a finite basis
expansion has the same effect as the introduction of an additional
operator in the hamiltonian. This is similar to the introduction of
constraining forces in classical mechanics, if the dynamics is
investigated under boundary conditions. Further, one can now
explicitely see that the ``coupled channel'' equations \xref{KS} and
the ``constrained'' forces \xref{Clas}, \xref{D} reduce to the standard KS-equations 
\xref{KS1} and Newton-equations \xref{Clas2}, respectively \xref{Clas1}, 
if the basis is complete.

\subsection{Energy and momentum balance}

In order to derive the energy balance we define an exchange-correlation energy according to
\beqx{
        \Axc[\rho] = \inttt\Exc[\rho](t) dt
}

with the important property
\beqx{
        \labex{del}
        \delfrac{\Axc[\rho]}{\rho(\r,t)} = \delfrac{\Exc[\rho](t)}{\rho(\r)} \quad .
}
Note the different arguments in $\delta\rho$ on the left and right hand side of \xref{del}.
With this, the potential energy of the quantum system can be written as
\beqx{ 
        \Epot(t)=\int\rho(\r, t)\left(V(\r, \R, t)+ \frac12\int\fracb{\rho(\r', t)} d^3r'\right)d^3r +\Exc[\rho](t).   
}%finish
The total time derivative of this functional is given by
\beqx{  
        \labex{ddtepot}
        \ddt{\Epot(t)}=\int d^3r\:\ddt\rho(\r,t)\cdot\Veff(\r, \R, t) + \int d^3r\;\rho(r,t)\cdot\ddt V(\r,\R,t)
}
with
\beqx{
        \labex{ddtrho}
        \ddt\rho(\r,t)=\pdt{}\rho(\r,t)+\sumA \dot\R_A\pdRA\rho(\r,t)
}
and
\beqx{
        \labex{ddtv}
        \ddt V(\r, \R, t) = \pdt{\VL(\r,t)} + \sumA \dot\R_A\pdRA\Vint(\r,\R) \quad .
}
Now, the total energy of the system can be defined
\beqx{ 
        E(t) = \sumA \frac{M_A}{2} \dot \R_A^2 + U(\R,t) + \sumjalbe\ajals\Talbe\ajbe 
        + \Epot[\rho](t)\quad .
} 
The total time derivative of this quantity is obtained after a longer calculation using
\xref{ddtepot}, \xref{ddtrho}, \xref{ddtv} and the equations of motion \xref{KS}  as
\beqx{
         \frac{dE}{dt}=\int\rho(\r,t)\pdt{\VL(\r,t)}d^3r-\sumA Z_A\pdt{\VL(\R_A, t)}  \quad .
}%finish
As expected, this quantity is conserved for vanishing or time-independent external fields.

A more transparent expression for the energy balance can be obtained
in dipole approximation (i.e. $\VL({\mathbf x},t)=-{\mathbf x} \cdot {\mathbf E}(t)$ ) leading to
\beqx{
        \ddt E = - \de(t)\dot\E(t)+\di(t)\dot\E(t)
}
with the dipole moments of the electrons
\beqx {
                \de(t) = \int \rho(\r,t)\r \; d^3r
}
and the ions
\beqx {
        \di(t) = \sumA Z_A \R_A(t) \quad .
}
From this expression it is clearly seen, that in a homonuclear system
($Z_A=Z=$const.) the ions will not be excited by the laser, because in
the center of mass system the nuclear dipole moment vanishes, i.e.
\beqx{
        \di=Z\sumA\R_A=0\quad.
}

In order to obtain more insight into the electronic excitation (deexcitation) process, it is convenient to consider the total energy change 
\begin{equation}
  \labex{eq:Echangetotal}
  \Delta E_{\text{el}} = - \int_{-\infty}^{\infty} \de(t)\dot{\E}(t)\, dt
\end{equation}
together with the Fourier-transformed dipole moment
\begin{equation}
  \labex{eq:dipoleft}
  \de(\omega) = \frac{1}{2\pi} \int_{-\infty}^{\infty} \text{e}^{-i\omega t} \de(t)\,dt \quad .
\end{equation}
One now immediately realizes that in a continuous wave field 
\begin{equation}
  \labex{eq:vecE}
  \E=\Re( \E_0 \, \text{e}^{-i\omega_{\text{L}} t})
\end{equation}
the electronic system adsorbs (desorbs) energy only if the imaginary part of 
$\de(\omega)$ does not vanish at the laser frequency $\omega=\omega_{\text{L}}$, i.e. 
\begin{equation}
  \labex{eq:Etotal3}
  \Delta E_{\text{el}} = \omega_{\text{L}}\, \Im( \de(\omega_{\text{L}}) \cdot \E_0) \quad .
\end{equation}
In the linear response region this is the case only if $\omega_{\text{L}}$ coincides with the excitation energy of an optical excited state.

In the other extreme case of very short laser pulses
\begin{equation}
  \labex{eq:vecEshort}
  \E=\E_0 \, \delta(t)
\end{equation}
all frequencies do contribute simultaneously to the excitation (deexcitation) process, i.e.
\begin{equation}
  \labex{eq:Eelshort}
  \Delta E_{\text{el}} = \int d\omega \, \omega\,\Im(\de(\omega) \cdot \E_0) \quad .
\end{equation}

For finite laser pulses, the total electronic energy change can be
obtained by solving \xref{eq:Echangetotal} numerically, together with
the the full equations of motion \xref{KS}, \xref{Clas} to calculate the dipole
moment $\de(t)$.

We note also, that the present formalism can be favourably used to
calculate optically excited states (i.e. Born-Oppenheimer surfaces) as well
as optical excitation spectra in the linear response region
from \xref{eq:dipoleft} by solving the KS-equations \xref{KS} for fixed
nuclear position $\R$ and "{}numerically short"{} $\delta$-pulses
\xref{eq:vecEshort}. Details of this procedure will be discussed
elsewhere \cite{tobe2}.

In order to investigate the momentum balance we start with the total momentum 
\beqx{ \P=\Pc+\Pq }%finish
as the sum of the classical part
\beqx{
         \Pc=\sumA M_A \dot \R_A 
}%finish
and the quantummechanical part
\beqx{
         \Pq=\sumj\bra{\Psi^j}{-i\nabla}{\Psi^j} = -i\sumjalbe\ajals\ajbe\braphi{\nabla} \quad. 
}%finish
Using the identity
\beqx{ 
        \parfraca{\r}\phial(\r-\RAal) = -\parfraca{\RAal}\phial(\r-\RAal)
        \equiv-\sumA \pdRA\phial(\r-\R_{A_\alpha})
}%finish
the latter one can be transformed into 
\beqx{
     \Pq=i\sumjalbe\ajals\sum_A\BAalbe\ajbe \quad .
}%finish
Now, the total derivative with respect to time can be obtained using \xref{KS} and \xref{Clas}
leading  after an extensive calculation to
\beqx{
        \labex{mom}
         \ddt \P = - \int \rho(\r,t) \nabla(\VL(\r, t)+\Vxc(\r,t))d^3r +\sum_A Z_A\nabla \VL(\R_A, t)\quad. 
}%finish
Besides the expected dependence on the laser field, this balance
contains a term that depends on the exchange-correlation potential $\Vxc \equiv
\delta \Axc/\delta\rho(\r,t)$. This one vanishes for the exact
$\Vxc$, which is a general property of TD-DFT \cite{Vignale95}. Without this
term one also immediately realizes, that in dipole approximation the
total momentum balance vanishes for neutral systems, i.e.
\begin{equation}
  \label{eq:mombal}
  \ddt \P = \left(-\int\rho(\r,t)d^3r +\sumA Z_A\right)\E(t) = 0
\end{equation}
which is due to the classical, not quantized treatment of the laser
field. 

We note finally that the momentum balance \xref{mom} can be derived
also (and much simpler) without basis expansion. The derivation,
carried out here, therefore proofs nicely the validity and stresses the importance
of the finite basis correction terms in the forces \xref{Clas}, \xref{D} following
from the variational principle.

\section{summary and outlook}

We have derived in a systematic way a generalized formalism of the
NA-QMD which applies for finite atomic many-body systems in external
fields. It is based on a mixed classical-quantum approach where the
electronic system is described by TD-DFT in local basis expansion and
the nuclear degrees of freedom are treated classically by
molecular dynamics. Self-consistent equations of motion
are derived from a general action principle.

We have presented here the exact equations of motion. They can be
solved without further approximations for one electron systems, like
\Hp\; or HD$^{+}$ \cite{tobe} where the exact exchange-correlation
term cancels the Hartree-term in the effective potential
\xref{Veff}. For many-electron systems, approximate equations of
motion, as derived e.g. in I on a tight-binding level, can be obtained
from the general formalism as well. We intend however, to realize the
numerical implementation of the whole formalism also on the ab-initio
level using the time-dependent optimized potential method
\cite{Gross98} for the exchange correlation part in the action
\xref{Apot1}. Preliminary results obtained within this method for
organic molecules, like ethylene C$_2$H$_4$, show excellent agreement
with CI-calculations \cite{Ben00} concerning the ground-state
properties (i.e. bonding lengths, angles etc.) as well as optical
excitation spectra \cite{tobe3}. As a first application of the whole
time-dependent formalism we intend to investigate the cis-trans
isomerization process of C$_2$H$_4$ in short laser pulses
\cite{tobe3}.

Another very interesting and fascinating field of application concerns
the excitation, ionization and fragmentation mechanism of atomic
clusters in intense laser fields
\cite{Ditmire96b,Ditmire97b,Pherson94,Lezius98,Ditmire99,Haberland98,Broer99}.
Here an all electron treatment together with an appropriate description of the continuum 
in the ansatz \xref{Basis} is required which, as discussed in the text, can be incorporated in the present formalism \cite{tobe2}.

This work was supported by the DFG through 
%Schwerpunkt ``Zeitabhängige Phänomene und Methoden in Quantensystemen der Physik 
%und Chemie'' as well as the 
Forschergruppe ``Nanostrukturierte Funktionselemente in makroskopischen Systemen''.

%\bibliography{all,theo} 

\end{document}